\documentclass[%
 reprint,
 superscriptaddress,
 amsmath,amssymb,
 aps,
]{revtex4-1}
\usepackage[colorlinks, citecolor=red]{hyperref}
\usepackage{graphicx}
\usepackage{dcolumn}
\usepackage{bm}

\usepackage{listings}
\usepackage{color}
\usepackage{subfigure}
\usepackage{algorithm}
\usepackage{bbm}

\usepackage{physics}
\usepackage[noend]{algpseudocode}
\usepackage{algorithmicx}
\usepackage{graphicx}

\usepackage{lipsum}

\definecolor{dkgreen}{rgb}{0,0.6,0}
\definecolor{gray}{rgb}{0.5,0.5,0.5}
\definecolor{mauve}{rgb}{0.58,0,0.82}

\makeatletter
\newenvironment{breakablealgorithm}
  {
   \begin{center}
     \refstepcounter{algorithm}
     \hrule height.8pt depth0pt \kern2pt
     \renewcommand{\caption}[2][\relax]{
       {\raggedright\textbf{\ALG@name~\thealgorithm} ##2\par}%
       \ifx\relax##1\relax 
         \addcontentsline{loa}{algorithm}{\protect\numberline{\thealgorithm}##2}%
       \else 
         \addcontentsline{loa}{algorithm}{\protect\numberline{\thealgorithm}##1}%
       \fi
       \kern2pt\hrule\kern2pt
     }
  }{
     \kern2pt\hrule\relax
   \end{center}
  }
\makeatother

\begin{document}
\bibliographystyle{apsrev4-1}
\title{Quantum Quantitative Trading: High-Frequency Statistical Arbitrage Algorithm}
\author{Xi-Ning Zhuang}
\affiliation{Key Laboratory of Quantum Information, CAS}
\affiliation{Origin Quantum Computing, Hefei, China}
\author{Zhao-Yun Chen}
\affiliation{Key Laboratory of Quantum Information, CAS}
\author{Yu-Chun Wu}
\email{wuyuchun@ustc.edu.cn}
\affiliation{Key Laboratory of Quantum Information, CAS}
\author{Guo-Ping Guo}
\email{gpguo@ustc.edu.cn}
\affiliation{Key Laboratory of Quantum Information, CAS}
\affiliation{Origin Quantum Computing, Hefei, China}

\begin{abstract}
  
Quantitative trading is an integral part of  financial markets with high calculation speed requirements, while no quantum algorithms have been introduced into this field yet. We propose quantum algorithms for high-frequency statistical arbitrage trading in this work by utilizing variable time condition number estimation and quantum linear regression. The algorithm complexity has been reduced from the classical benchmark $O(N^2d)$ to $O(\sqrt{d}\kappa_0^2\log{(1/\epsilon)}^2))$. It shows quantum advantage, where $N$ is the length of trading data, and $d$ is the number of stocks, $\kappa_0$ is the condition number and $\epsilon$ is the desired precision. Moreover, two tool algorithms for condition number estimation and cointegration test are developed.

\end{abstract}
\maketitle

\section{INTRODUCTION}

With the rapid development of quantum computing\cite{divincenzo2000physical,kalai2011quantum,arute2019quantum}, the qubits on the chips are up to 53 currently\cite{arute2019quantum}, and it will extend beyond  100 soon in the roadmap of quantum systems based on superconductivity. Hence, quantum computing shows the potential to solving practical problems, such as chemistry\cite{mcardle2020quantum,outeiral2021prospects,emani2021quantum}, materials\cite{ma2020quantum}, drug design\cite{cao2018potential}, and et al.

Quantum computation has produced positive effects in finance\cite{orus2019quantum,egger2020quantum}, and current quantum algorithms mainly focus on solving derivatives pricing problems and risk analysis by quantum Monte-Carlo(QMC) simulation\cite{ceperley1986quantum,montanaro2015quantum,stamatopoulos2020option,martin2019towards,rebentrost2018quantum,woerner2019quantum},  optimizing stocks portfolio through quadratic unstrained binary optimization(QUBO)\cite{rosenberg2016solving,lopez2015generalized,rosenberg2016finding}, and financial analysis work utilizing quantum machine learning(QML)\cite{wittek2014quantum,schuld2015introduction,lloyd2013quantum,buhrman2001quantum}. However, for quantitative trading and especially statistical arbitrage, there are no corresponding quantum algorithms yet.

Quantitative trading is an essential field of finance, and statistical arbitrage is a mainstream approach of quantitative trading taken by most hedge funds \cite{krauss2017statistical,kanamura2008application}. While lots of classical algorithms for quantitative trading have been proposed \cite{gatev2006pairs,vidyamurthy2004pairs,caldeira2013selection,elliott2005pairs}, and traditional hardware techniques including infrared communication and Field Programmable Gate Array have been employed over the years \cite{leber2011high,narang2013inside}, still the requirement for speed cannot be satisfied when implementing those complicated statistical methods, especially in the quicker-take-all situation of high-frequency trading(HFT) whose need of computing speed is crucial\cite{gomber2015high}.  In statistical arbitrage, one needs to find a potential cointegrated pair via many linear regressions and cointegration tests involving a huge matrix of historical data. For example, in U.S. stock markets, the problem size can exceed  $N=10^7$ and the complexity is $10^{15}$(see section \uppercase\expandafter{\romannumeral6} for details) which is very hard to calculate by classical computers.  For this problem, quantum computation might provide an effective solution. 

In this article, quantum algorithms applied to statistical arbitrage strategy are proposed. It consists of two subroutines: the first one is the Variable Time Preselection Algorithm(VTPA) that will help to find, with high probability, the potential comovement out of  securities and portfolios. The second one is the Quantum Cointegration Test Algorithm(QCTA) that focuses on the efficient verification of cointegrated pairs, which is quite valuable in statistics but has not been achieved via quantum computation ever before. The classical benchmark to achieve the preselection procedure is by matrix factorization with complexity $O(N^3)$ \cite{trefethen1997numerical},  while our algorithm's complexity is $O(\sqrt{d}\kappa_0^2\log{(1/\epsilon)}^2))$ where $d$ is the number of stocks usually much less than time length $N$ and $\kappa_0$ is the condition number. Moreover, an efficient tool named Quantum Condition Number Comparison Algorithm (QCNCA) used to probe a matrix's condition number is proposed, and it can be applied to many other domains.

The structure of this article is as follows:  After giving the preliminaries in section \uppercase\expandafter{\romannumeral2}, the global structure and main results of our work are shown in section \uppercase\expandafter{\romannumeral3}.  The details of VTPA and QCT are described in sections \uppercase\expandafter{\romannumeral4} and \uppercase\expandafter{\romannumeral5}, respectively, followed by a discussion on complexity and quantum advantage in section \uppercase\expandafter{\romannumeral6}.

\section{PRELIMINARIES}
Since different domains, including quantum computing, statistics, and finance, are covered while the readers may not be familiar with one of them, related preliminaries are introduced in detail. 

\subsection*{A. multicollinearity}
In this subsection, a brief introduction about multicollinearity and condition number will be given, helping to understand the first preselection algorithm. In statistics, \textbf{multicollinearity} refers to a situation in which some of the explanatory variables in a multiple regression model are highly linearly related. 
 
In numerical analysis, to detect and measure the seriousness of the multicollinearity problem, the \textbf{condition number $\kappa$} is introduced.  Given problem $f$, it is generally a measurement to describe the change of the output value divided by the change of the input variable x:
\[\kappa(f)=\lim_{\epsilon\to0}\sup_{\lVert\delta x\rVert\le\epsilon}\frac{\lVert\delta f\rVert}{\lVert\delta x\rVert}. \]
In the case of matrices, the condition number associated with the linear equation $Ax=b$ release the dependence of accuracy on the input data.  Specifically, the condition number of normal matrix  $A$ is
\[\kappa(A) = \frac{\abs{\lambda_{max}(A)}}{\abs{\lambda_{min}(A)}}. \]
It should be emphasized that condition number is a property of the matrix itself and does not depend on the algorithm or accuracy of the computer used.  Hence, both classical and quantum computers have a common problem to solve an ill-conditioned (high condition number) linear system of equations.  The larger the condition number, the more ill-conditioned the matrix is, and the algorithm complexity will increase very quickly.

\subsection*{B. Cointegration}
In this subsection, some statistical concepts and facts about stochastic process and time series analysis are provided.  Following those, an explicit demonstration is also given on the relationship between multicollinearity and cointegration, which may be confuse some readers.
 
A (weakly) \textbf{stationary} time series, $x_t$, is a finite variance process with an unconditional joint probability distribution. Thus it does not change when shifted in time: (i) the mean value function $\mu_{xt} = E(x_t)$ is constant and (ii) the covariance function $\gamma_{x}(s, t) = E[(x_s-\mu_s)(x_t-\mu_t)]$ depends on s and t only through their difference$\abs{s-t}$. In autoregressive-moving average models of unknown order, to test whether a given time series denoted as $Y_t$ is stationary or not, the Augmented Dickey-Fuller(ADF) unit root test  may be employed\cite{said1984testing}.

\textbf{Cointegration}  (multi-cointegration) is a relevant statistical property of two or more time series which are individually integrated of order d while their combination is integrated of order less than d. Here the \textbf{order of integration} is a summary statistic denoting the minimum number of differences taken to obtain a covariance-stationary series. Without loss of generality, $d=1$ is assumed in this article. Under different financial hypotheses, there are mainly three kinds of cointegration tests: the Engle–Granger test\cite{engle1987co}, the Johansen test\cite{johansen1990maximum}, and the Phillips-Ouliaris test \cite{phillips1988testing}.  In our work, Engle–Granger two-step method is used as the most popular and famous one:  

Suppose that $x^i_{t}$ are non-stationary and integrated of order d=1, then a linear combination 
\[{\hat {u}}_{t} = \sum{\beta_ix^i_t}\]
is expected to be stationary for some specific coefficient of $\beta_i$. In the general case that $\beta_i$ is not decided yet, some estimation must be made first, usually by ordinary least squares regression.  Next, the stationarity test will be implemented on the residuals ${\hat {u}}_{t}$. It is a regression on ${\hat {u}}_{t}$, and the lagged residuals  $\hat{u}_{t-1}$ are included as a regressor:
\[\Delta u_t = \alpha + \beta t + \gamma u_{t-1} + \sum_{i=1}^{p-1}\Delta u_{t-i} + \epsilon_t. \]
Here $\alpha$ and $\beta$ are the intercept and the coefficient on the time trend, respectively, and $p$ denotes the lag order of the autoregressive process to be decided. The unit root test is carried out under the null hypothesis $\gamma=0$. The test statistic to be computed is 
\[\mathbf{DF_\tau}=\frac{\hat{\gamma}}{SE(\hat{\gamma})}.\]
 What follows is a comparison with the Dickey-Fuller distribution critical value table \cite{fuller1976introduction}. 

Whenever such a cointegrated stock portfolio is found, the linear combination is expected to have the property of mean-reverting and use the statistical arbitrage.

\subsection*{C. Quantum Linear Regression}

    Quantum linear regression is the primary tool of QCT and is introdecd as follows. Wiebe, Braun, and Lloyd (WBL)firstly introduced an algorithm for quantum data fitting\cite{wiebe2012quantum}.  Building on Harrow, Hassidim, and Lloyd’s (HHL) quantum algorithm for linear systems of equations\cite{harrow2009quantum}, WBL developed a least-squares estimation using Moore–Penrose Pseudo inverse.  WBL’s algorithms are mainly suited for data sets whose design matrices are sparse and well-conditioned.  Given an $N$ dimension $s$ sparse data matrix, the time complexity is $O(\log N s^3 \kappa^6\epsilon^{-1})$, where the condition number given is $\kappa$ and the accuracy desired is $\epsilon^{-1}$.  With the technique of quantum principal component analysis(qPCA) and singular value decomposition(SVD) \cite{lloyd2014quantum}, Schuld, Sinayskiy, and Petruccione(SSP) came with an algorithm for prediction based on  a linear regression model with least-squares optimization \cite{schuld2016prediction}.  The sparseness condition is removed, and the existence of a low-rank approximation is supposed instead.  The time complexity is $O(\log N \kappa^2\epsilon^{-3})$, where an improvement of factor $\kappa^4$ is made on the condition number at the cost of worse dependence on accuracy by a factor $\epsilon^{-2}$.  Recently, Guoming Wang presents a new quantum algorithm for fitting a linear regression model using least-squares approach \cite{wang2017quantum}.  This algorithm builds on Low and Chuang’s method for Hamiltonian simulation based on qubitization and quantum signal processing \cite{low2017optimal,low2019hamiltonian}. Childs, Kothari, and Somma (CKS)'s approach is introduced to inverse the matrix derived from SVD \cite{childs2017quantum}. Imposing restrictions on the number of adjustable parameters $d$, and hence the rank of the design matrix, the gate complexity is $O(\frac{d^{1. 5}\kappa^3}{\epsilon^2}{\rm poly}[\log_2(\frac{\kappa}{\epsilon\delta})])$ with the succeeding probability is at least $1-\epsilon$. 

\section{QUANTUM STATISTICAL ARBITRAGE}

Pioneered by Gerry Bamberger \cite{bookstaber2007demon}, statistical arbitrage has developed a lot, and the crux and core are to model the comovement. Following the framework first introduced by Vidyamurthy\cite{vidyamurthy2004pairs}, statistical arbitrage is divided mainly into three key steps:  Firstly, two or more securities moved together historically in a formation period should be preselected;  secondly, some version of the Engle-Granger cointegration test\cite{engle1987co} is taken for verification; thirdly, the spread between them in a subsequent trading period is  monitored by some optimal entry/exit thresholds. Since the spread of stocks will revert to its historical mean and, the profit can be made from other traders' irrational behavior by longing the oversold securities and shorting the overbought ones at the same time\cite{gatev2006pairs}. 

In this section, two algorithms solving the quantum statistical arbitrage problem are proposed. One is for the case of fixed condition number threshold; the other is for a fixed number of remained portfolios. The formal statement of the quantum statistical arbitrage problem is as follows: Given historical data of many stocks for a long time interval, our target is to select those stocks that are cointegrated. The algorithm mainly contains two steps: preselect multicollinear stock portfolios from the pool by applying $VTPA(p, \kappa)$ where $VPTA$ is True if the given portfolio $p$'s condition number is larger than the threshold $\kappa$; and then verify whether the preselected portfolio $p$ is cointegrated by implementing $QCT(p)$ to output cointegration flag $f$ and corresponding coefficients $\beta$.

Suppose that $P=\{p\}$ is the portfolio pool, and $(p_t^{(j)})_{J\times T}$ is a portfolio of stocks' historical quote data. Here ${p_t^{(j)}}$ is an element of $p$ as the $j^{th}$ stock's price at time t. The matrix $p$  is of full rank since no perfect linear relation exists in noisy financial market data. The two quantum statistical algorithms work in the standard oracle model, and the matrix is stored in a quantum random access memory(qRAM) \cite{giovannetti2008quantum, giovannetti2008architectures, hong2012robust}. A procedure $\mathcal{P}_x$ is assumed to perform the map
\[\ket{j}\ket{t}\ket{z}\rightarrow\ket{j}\ket{t}\ket{z\oplus p_t^{(j)}}\]
for any $j\in[1,2,...,d]$ and $t\in[1,2,...,N]$, and the price is stored as a bit string in the third register.

In order to derive the desired real symmetric matrix, the strategy of HHL \cite{harrow2009quantum,wiebe2012quantum} is adopted  as:
\[A=\bigl(\begin{smallmatrix}
0&X\\x^T&0
\end{smallmatrix} \bigr).\]
Moreover, the norm of the matrix is assumed to satisfy $\lVert A \rVert=1$ without loss of generality since otherwise let $A=\frac{A}{\lVert A \rVert}$.

If an efficient $\kappa_0$ derived from historical data is taken as filter threshold, the following \textbf{Algorithm} \ref{alg:alg1} is given:\\ 

\begin{breakablealgorithm}
\caption{Quantum Statistical Arbitrage Algorithm with Fixed Condition Number Preselection}
\label{alg:alg1}
\begin{algorithmic}[]
\Require ~~\\
$\kappa_0$:  the threshold for preselection\\
$T$:	the length of time interval\\
$J$:	the total number of stocks\\
$d$: number of stocks in one portfolio\\
$P$: the portfolio pool set contains portfolios p\\
$p_t^{(j)}$: the $j^{th}$ stock's price at time t.
\Ensure ~~\\
$(p,\beta)$ Cointegrated portfolios and cointegration coefficients.\\

\State Data Loading
\For{$p$ in $P$}
\State $\ket{p}=\sum\limits_{t=0}^{T-1}\sum\limits_{j=0}^{J-1}p_t^{(j)}\ket{t}\ket{j}$
\If{$VTPA(p,\kappa_0) = True$}
\State $QCT(p) = f,\beta$
\If{$f=True$}
\State Output $(p,\beta)$
\EndIf
\Else
\State Skip to the next loop
\EndIf
\EndFor
\end{algorithmic}
\end{breakablealgorithm}
\noindent \\

As for the case of unknown $\kappa_0$, an even more efficient \textbf{Algorithm} \ref{alg:alg2} is provided. The basic idea is as follows: since our single-step preselection sub-algorithm can be used for any given $\kappa$, a progressive $\kappa$ preselection procedure can be implemented. Portfolio matrices with small $\kappa$ will be directly obsoleted in the first several steps until the number of matrices left is small enough, and until then, the quantum cointegration test will be implemented.\\

\begin{breakablealgorithm}
\caption{Quantum Statistical Arbitrage Algorithm with Progressive Preselection}
\label{alg:alg2}
\begin{algorithmic}[]
\Require ~~\\
$k$: portfolio number threshold\\
$T$:	the length of time interval\\
$J$:	the total number of stocks\\
$d$: the number of stocks in one portfolio\\
$P$: the portfolio pool\\
$p_t^{(j)}$: the $j^{th}$ stock's price at time t.
\Ensure ~~\\
$(p,\beta)$ Cointegrated portfolios and cointegration coefficients.\\

\State Data Loading.
\State Step counter $j = 1$
\State Portfolio counter $K=\abs{P}$
\While{$K>k$}
\State{$\kappa_j=2^j$}
\For{$p$ in $P$}
\State $\ket{p}=\sum\limits_{t=0}^{T-1}\sum\limits_{j=0}^{J-1}p_t^{(j)}\ket{t}\ket{j}$
\If{$VTPA(p,\kappa_j) = True$}
\State skip
\Else
\State $K = K - 1$
\State $P=P-\{p\}$
\EndIf
\EndFor
\State $j = j + 1$
\EndWhile
\For{$p$ in $P$}
\State $QCT(p) = (f,\beta)$
\If{$f=True$}
\State Output $(p,\beta)$
\EndIf
\EndFor
\end{algorithmic}
\end{breakablealgorithm}
\noindent \\
Both of the above two algorithms are for statistical arbitrage, and the selection depends on the specific market: if the $\kappa$-threshold is stationary, the first algorithm is chosen; otherwise, the second one is preferred.
Since the two subroutines are complicated and tool sub-algorithms are developed, they will be introduced in section \uppercase\expandafter{\romannumeral4} and section \uppercase\expandafter{\romannumeral5}, respectively.

\section{VARIABLE TIME PRESELECTION}

In this section, we will explain the main idea of the first part of our work as a variable time quantum algorithm to preselect the stocks that are multicollinear and thus may be cointegrated as needed.

Although ill-conditioned matrices are commonly considered a terrible problem that one should try to avoid, we develop the heuristic idea to detect multicollinearity by searching matrices with small eigenvalues and large condition numbers.  QCNCA is developed to determine whether the condition number $\kappa$ of a given matrix is larger than the threshold $\kappa_0$ in subsection A. 

Since QCNCA's dependence on $\kappa$ is quadratic, the technique of variable time quantum algorithm is introduced to accelerate the implementation of matrices selection\cite{ambainis2012variable}, and then the VTPA is as follows:\\

\noindent\textbf{Theorem 1} Supposing that many different linear systems are given with unknown condition number $\kappa$ and $P_j$ denote the probability that condition number satisfies $\kappa_{j-1}=2^{j-1}\le\kappa\le\kappa_j=2^j$.  Then there is an efficient quantum algorithm to preselect matrices with condition numbers $\kappa\ge\kappa_0$.  The average query complexity is $O(\sqrt{d}\log{(1/\epsilon)}^2(\sum_{j=1}^M 4^jjP_j))$.  As for a uniform probability distribution, the query complexity is $O(\sqrt{d}\kappa_0^2\log{(1/\epsilon)}^2)$ to determine whether the condition number is larger than $\kappa_0$. \\

The proofs of correctness and complexity of \textbf{Theorem 1} are given in subsection C and subsection D, respectively.

\subsection*{A. Tools: Quantum Condition Number Comparison Algorithm}

Realizing that multicollinearity appears with large $\kappa$ \cite{pesaran2015time,belsley2005regression}, and hence small eigenvalues, the following preselection algorithm is developed: repeat a simplified  phase estimation sub-algorithm until an eigenvalue small enough is detected.  If such an eigenvalue is found, the corresponding portfolios will be recorded as an alternative one.  It worth noticing that some cointegrated pairs may be missed in our algorithm, but it does not matter since our task is to search for some collinear portfolios instead of the impossible mission to find all of the cointegrated pairs.  We denote this procedure Quantum Condition Number Comparator $QCNC(\kappa, \varphi)$ and get the following result:\\

\noindent\textbf{Lemma 2} Supposing that $A$ is an $N\times N$ Hermitian matrix with $\lVert A \rVert = 1$ with unknown condition number $\kappa$ and the probability density function of eigenvalues is $p(\lambda)$.  Then there is a quantum algorithm using $O(\kappa_0\log{(1/\epsilon)\frac{\int_{1/\kappa}^1 p(x)\, dx}{\int_{1/\kappa_0}^{1/\kappa} p(x)\, dx}})$ calls of A to determine whether the condition number is larger than $\kappa_0$.  In the case of a uniform probability distribution, A's calls are $O(\kappa_0^2\log{(1/\epsilon)})$ so that whenever $\kappa\ge2\kappa_0$, the target qubit will be 1. \\

It should be noticed that this repeating time, especially when $\kappa$ is large, is determined by the threshold $\kappa_0$, while traditional algorithms depend on the unknown $\kappa$. This is an algorithm finding whether the condition number of a linear system is large than the given threshold without solving the equations.\\

\noindent$Proof\ of\ \textbf{Lemma 2}. $ Without loss of  generality, suppose that $A$ is a matrix with Frobenius norm
\begin{equation}
\lVert A \rVert_F=(\sum_{i=1}^m\sum_{j=1}^n\abs{a_{ij}}^2)^{1/2}=\sqrt{d}
\end{equation}
 (otherwise let $A=\frac{\sqrt{d}}{\lVert A \rVert_F}A$), and unknown rank $r$. A direct calculation shows that:
\begin{align}
\abs{\lambda_{max}(A)}=&\lVert A \rVert_2\\
\ge&\frac{1}{\sqrt{r}}\lVert A \rVert_F\\
=&\sqrt{d/r}\\
\ge&1.
\end{align}
Here in (2)
\begin{equation}
\lVert A \rVert_2=\sup\limits_{x\ne0}{\frac{\lVert Ax \rVert}{\lVert x \rVert}}=\sigma_{max}(A)
\end{equation}
is the induced $L_2$ norm and equals to $\abs{\lambda_{max}(A)}$ (see \cite{horn1985johnson}'s example 5.6.6), and it follows the inequality (3) (see \cite{golub1996cf}).

For any given eigenvector $\ket{\lambda}$, with a variant of phase estimation, it is easy for us to determine whether its corresponding eigenvalue $\lambda$ is larger than $1/\kappa_0$ or not with complexity $O(\kappa_0 \log(1/\epsilon))$ \cite{childs2017quantum}. By the definition of the condition number of normal matrices, for any known eigenvalue $\lambda$:
\begin{equation}
\kappa=\frac{\abs{\lambda_{max}(A)}}{\abs{\lambda_{min}(A)}}\ge\frac{1}{\abs{\lambda(A)}}\ge\kappa_0.
\end{equation}
Hence a lower bound of the condition number is also given.  Whenever a sufficiently small eigenvalue $\lambda_0$ is given, the matrix can be regarded with condition number greater than $\kappa_0$ and high multicollinearity as a consequence. 

Obviously, there is a certain probability of success when the testing eigenvalue is larger than $\kappa_0$.  Let the condition number be $\kappa$ and the probability density function of eigenvalues be $p$; the success probability is
\begin{equation}
P_{success}=\frac{\int_{1/\kappa_0}^{1/\kappa} p(x)\, dx}{\int_{1/\kappa}^1 p(x)\, dx}.
\end{equation}
Under the assumption that the eigenvalues follow a uniform probability distribution, the success probability turns to be
\begin{align}
P_{success}=&\frac{1/\kappa_0-1/\kappa}{1-1/\kappa}\\
=&\frac{1}{\kappa_0}\frac{\kappa-\kappa_0}{\kappa-1}\\
\approx&\frac{1}{\kappa_0}(1-\frac{\kappa_0}{\kappa}).
\end{align}
Here $\kappa$ is assumed large and $\kappa-1 \simeq \kappa$.  Moreover, whenever a matrix with $\kappa\ge2\kappa_0$ is given, we have:
\begin{equation}
P_{success}\ge 1/2\kappa_0.
\end{equation}
This procedure shall be repeated $2\kappa_0$ times to boost the success probability.  Hence the total number of calls for A is $O(\kappa_0^2\log{(1/\epsilon)})$. Since complexity to simulate $U=e^{iA}$ is $O(\sqrt{d}(1+\log{(\kappa_0/\epsilon)}))$\cite{berry2015hamiltonian}, the total query complexity is $O(\sqrt{d}\kappa_0^2\log{(1/\epsilon)}(1+\log{(\kappa_0/\epsilon)}))$. $\hfill\blacksquare$
\\[1pt]

It should be mentioned that the assumption of uniform distribution is reasonable. Although different distributions of eigenvalues may appear in specified realistic problems, some normal conditions can be imposed to guarantee that the algorithm will still work with slight modification. 

\subsection*{B. Algorithm}

To see how to derive an algorithm more efficient on $\kappa$, one should notice that matrices with small condition numbers can be found quite early and need not be calculated anymore. Some clock registers are used to obsolete those matrices with small condition numbers by starting from small threshold $\kappa_0$.  The larger the threshold $\kappa_0$ is, the fewer the matrices need to be tested.  Repeating this procedure several times can make the acceleration.

Suppose that $M=\lceil\log{\kappa_0}\rceil$.  The clock registers $C_{1, . . . , M}$ are used to control and store the result of subprocedure $\mathcal{A}_j$ defined later.  Another 1-qubit register  $\mathcal{F}$ is used as a flag register to donate if the algorithm is stopped.  For all $j \in \{1,. . . ,M\}$, let $\phi_j=1/\kappa_j = 2^{-j}$, and let $\epsilon$ be the desired precision.  Since it is our target to verify whether matrix $A$ contains components corresponding to small eigenvalues, the algorithm is defined as $\mathcal{A}=\mathcal{A}_M\mathcal{A}_{M-1}. . . \mathcal{A}_1$, where $\mathcal{A}_j$ is defined as follows:\\

\noindent\textbf{Algorithm $\mathcal{A}_j$} Conditional on first $j-1$ qubits of $\mathcal{H_C}$ being $\ket{1}$, apply QCNC($\kappa_j, \epsilon$) using $C_j$ as the output qubit and additional fresh qubits from $\mathcal{P}$ as ancilla (denoted by $P_j$). If $C_j$ is left $\ket{0}$ in the first term, the qubit on stop flag register $\mathcal{F}$ will be flipped.

\subsection*{C. Correctness}
We shall now prove the correctness of this algorithm. \\

\noindent$Proof\ of\ \textbf{Theorem 1 (correctness part)}. $ Given a matrix A, the condition number is either in some interval $[\kappa_j, 2\kappa_{j}]$ or greater than $\kappa_0$.\\

\noindent i) Suppose that a matrix with condition number $\kappa_j \le \kappa \le 2\kappa_j$ is given:\\

\noindent\textbf{State after $\mathcal{A}_1$ to $\mathcal{A}_{j-1}$} 
Since $\kappa \le \phi_{1, . . . , j-1}$, the clock registers $C_1, . . . , C_{j-1}$ are at position $\ket{1}$ while the stop flag register $\mathcal{F}$ stays $\ket{0}$ with high probability.  After j-1 steps the state is left as
\[\ket{1}_{C_1,. . . ,C_{j-1}}\ket{0}_{C_j,. . . ,C_M}\ket{0}_\mathcal{F}\]
\[\ket{\gamma^1_1}_{P_1}. . . \ket{\gamma^{j-1}_1}_{P_{j-1}}\ket{0}_{P_j. . . P_M}\] 
where $\ket{\gamma^{i}_1}$ is the ancillary state produced by the $i^{th}$ call to QCNC.\\

\noindent\textbf{State after $\mathcal{A}_j$} 
Because $\phi_j \le \kappa \le 2\phi_j$, QCNC will split the $j^{th}$ control register to $\ket{1}_{C_j}$ with high probability:
\begin{align*}
\beta_0\ket{\mathcal{U}_{j-1}}_\mathcal{C}\ket{0}_\mathcal{F}\ket{\gamma^1_1}_{P_1}. . . \ket{\gamma^{j-1}_1}_{P_{j-1}}\ket{\gamma^j_0}_{P_j}\ket{0}_{P_{j+1}. . . P_M}&\\
+\beta_1\ket{\mathcal{U}_{j}}_\mathcal{C}\ket{0}_\mathcal{F}\ket{\gamma^1_1}_{P_1}. . . \ket{\gamma^{j-1}_1}_{P_{j-1}}\ket{\gamma^j_1}_{P_j}\ket{0}_{P_{j+1}. . . P_M}&\\
\text{where } \mathcal{U}_j = 1^{j}0^{m-j}. &
\end{align*}
Since $C_j$ is left $\ket{0}$ in the first term, the qubit on register $\mathcal{F}$ is flipped:
\begin{align*}
\beta_0\ket{\mathcal{U}_{j-1}}_\mathcal{C}\ket{1}_\mathcal{F}\ket{\gamma^1_1}_{P_1}. . . \ket{\gamma^{j-1}_1}_{P_{j-1}}\ket{\gamma^j_0}_{P_j}\ket{0}_{P_{j+1}. . . P_M}&\\
+\beta_1\ket{\mathcal{U}_{j}}_\mathcal{C}\ket{0}_\mathcal{F}\ket{\gamma^1_1}_{P_1}. . . \ket{\gamma^{j-1}_1}_{P_{j-1}}\ket{\gamma^j_1}_{P_j}\ket{0}_{P_{j+1}. . . P_M}. &
\end{align*}

\noindent\textbf{State after $\mathcal{A}_{j+1}$} This will affect the two parts of the state in different way:
In the case that the $j^{th}$ control qubit is splitted, the step QCNC($\kappa_{j+1}, \epsilon$) is implemented.  Notice that $\kappa \le \phi_{j+1}$, the state turns to be:
\begin{align*}
\beta_1\ket{\mathcal{U}_{j}}_\mathcal{C}\ket{1}_\mathcal{F}\ket{\gamma^1_1}_{P_1}. . . \ket{\gamma^j_1}_{P_j}\ket{\gamma^{j+1}_0}_{P_{j+1}}\ket{0}_{P_{j+2}. . . P_M}. 
\end{align*}
As for the case that $j^{th}$ control qubit is $\ket{0}$, nothing will be done and the state is:
\begin{align*}
\beta_0\ket{\mathcal{U}_{j-1}}_\mathcal{C}\ket{1}_\mathcal{F}\ket{\gamma^1_1. . . \gamma^{j-1}_1}_{P_1. . . P_{j-1}}\ket{\gamma^j_0}_{P_j}\ket{0}_{P_{j+1}. . . P_M}&\\
+\beta_1\ket{\mathcal{U}_{j}}_\mathcal{C}\ket{1}_\mathcal{F}\ket{\gamma^1_1. . . \gamma^{j}_1}_{P_1. . . P_{j}}\ket{\gamma^{j+1}_0}_{P_{j+1}}\ket{0}_{P_{j+2}. . . P_M}&. 
\end{align*}\\

\noindent\textbf{State after $\mathcal{A}$}

Given a matrix $A$ with condition number $\kappa_j\le\kappa<\kappa_{j+1}$, the final state at the end of algorithm $\mathcal{A}$ is:
\begin{align*}
\beta_0\ket{\mathcal{U}_{j-1}}_\mathcal{C}\ket{1}_\mathcal{F}\ket{\gamma^1_1. . . \gamma^{j-1}_1}_{P_1. . . P_{j-1}}\ket{\gamma^j_0}_{P_j}\ket{0}_{P_{j+1}. . . P_M}&\\
+\beta_1\ket{\mathcal{U}_{j}}_\mathcal{C}\ket{1}_\mathcal{F}\ket{\gamma^1_1. . . \gamma^{j}_1}_{P_1. . . P_{j}}\ket{\gamma^{j+1}_0}_{P_{j+1}}\ket{0}_{P_{j+2}. . . P_M}&. 
\end{align*}

ii)As for matrix $A$ with condition number $\kappa\ge\kappa_0$, we have:
\[\ket{1}_\mathcal{C}\ket{0}_\mathcal{F}\ket{\gamma^1_1. . . \gamma^{M}_1}_{P_1. . . P_{M}}\]
It should be noticed that whenever the flag register is splitted to $\ket{1}_\mathcal{F}$, the algorithm stops at some step and reject the hypothesis of the condition number larger than $\kappa_0$.  Hence a measurement can be implemented on $\ket{1}_F$ to decide whether the matrix contains a component with eigenvalues less than some given $1/\kappa_0$.  Besides, a control counter circuit can be employed on the M clock registers to probe the range for $\kappa$. 

\subsection*{D. Complexity Analysis}
In this subsection, the algorithm's complexity analysis is given to finish the proof of \textbf{Theorem 2}.  It should be mentioned that the algorithm complexity depends on the specific distribution of condition numbers and eigenvalues of the problem.  In this work, a theoretical framework is developed for analysis.  Moreover, as an example, the result assuming that $\log{\kappa}$ follows a uniform probability distribution is calculated.  This assumption is common and reasonable since there are relatively fewer matrices with large condition number.\\

\noindent$Proof\ of\ \textbf{Theorem 1 (complexity part)}. $
Suppose that there are n matrices and the condition number threshold for comparison is $\kappa_0$.  Let $M=\lceil \log{\kappa_0} \rceil$.  Let $\kappa_j=2^j$ and $P_j$ be the probability that the matrix's condition number satisfies $\kappa_{j-1}\le\kappa\le\kappa_j$.  Then the cummulative number of queries $T_j$ for this kind of matrix is:
\begin{align}
T_j&=\sum_{k=1}^j \text{QCNC}(\kappa_k, \epsilon)\\
&=\sum_{k=1}^j \kappa_k^2 \log{(1/\epsilon)}\sqrt{d}(1+\log{(\kappa_k/\epsilon)})\\
&=\sqrt{d} \log{(1/\epsilon)}^2 \sum_{k=1}^j 2^{2k+1}k\\
&=\sqrt{d} \log{(1/\epsilon)}^2 \frac{(j-1/3)4^{j+1}+4/3}{3}\\
&\le\frac{4^{j+1}j}{3} \sqrt{d} \log{(1/\epsilon)}^2. 
\end{align}
Hence cnosidering the probability, the arithmatic average number of queries is:
\begin{align}
T_{avg}=&\sum_{j=1}^M P_jT_j\\
\le&\sum_{j=1}^M \frac{4^{j+1}j}{3} \sqrt{d}\log{(1/\epsilon)}^2 P_j\\
=&\frac{4}{3}\sqrt{d}\log{(1/\epsilon)}^2(\sum_{j=1}^M 4^jjP_j). 
\end{align}
Supposing that $\log{\kappa}$ follows a uniform probability contribution, the probability is 
\[P_j=1/M=1/\log{\kappa_0}\]
and the average time is:
\begin{align}
T_{avg}&\le\frac{4}{3}\sqrt{d}\log{(1/\epsilon)}^2(\sum_{j=1}^M 4^jjP_j)\\
&=\frac{4}{3}\sqrt{d}\log{(1/\epsilon)}^2(\sum_{j=1}^M 4^jj/M)\\
&=\frac{4}{3M}\sqrt{d}\log{(1/\epsilon)}^2(\sum_{j=1}^M 4^jj)\\
&=\frac{4}{3M}\sqrt{d}\log{(1/\epsilon)}^2\frac{M4^{M+1}}{3}\\
&\le\frac{16}{9}\sqrt{d}\log{(1/\epsilon)}^2 4^M\\
&=\frac{16}{9}\sqrt{d}\kappa_0^2\log{(1/\epsilon)}^2
\end{align}
Hence the complexity is $O(\sqrt{d}\kappa_0^2\log{(1/\epsilon)}^2)$ as claimed in \textbf{Theorem 1}. 
\hfill$\blacksquare$
\\[8pt]
The complexity of this algorithm also depends on the fixed threshold $\kappa_0$ instead of an  unknown $\kappa$.  Hence besides the acceleration compared to classical algorithms, the stability and robustness are also improved to satisfy financial problems. 

\section{QUANTUM COINTEGRATION TEST}

To finish the last peice of quantum statistical arbitrage, it needs to be verified whether the preselected matrices contain a cointegrated pair. The global structure and details of QCT are described in the first subsection, and the analysis of complexity is given in the second subsection. These two parts yield the following result:\\
\noindent \textbf{Theorem 3} Suppose that $d$ and $N$ are the number of kinds of stocks and the time length of stock prices, $\epsilon$ is the precision desired, and $\kappa$ is the condition number. Then the cointegration test with $L$ lag-length  augmented dickey fuller test can be implemented with complexity $O(\frac{d^{2. 5}\kappa^3}{\delta^2}{\rm poly}(\log_2{\frac{d \kappa}{\delta}})+dN+\frac{{(L+2)}^{2. 5}\kappa'^3}{\delta'^2}{\rm poly}(\log_2{\frac{(L+2)\kappa'}{\delta'}})$, where $\delta=min\{1/d,\epsilon\}$, $\delta'=min\{1/(L+2),\sqrt{L+2}\epsilon^2\}$.

\subsection*{A. Algorithm}
First of all, the following procedure is used to generate the residual sequence of linear regression. Since the residuals sequence is needed instead of regression coefficients or predicted values\cite{wang2017quantum, schuld2016prediction}, known quantum linear algorithms should be employed with some further modification. The work of \cite{wang2017quantum}'s Theorem2 is used to derive an approximation $\beta$ of the regression coefficients $\hat{\beta}$. \\ 

\noindent \textbf{Lemma 4} (QLR, Theorem 2 in \cite{wang2017quantum}) Let $\textbf{X}=(x_{i,j})$ be an $N*d$ balanced matrix such that its singular values are in range $[1/\kappa, 1]$. Let $\textbf{y}={(y_1,y_2,...,y_N)}^T$ be a balanced unit vector. Suppose $(\textbf{X}, \textbf{y})$ is well behaved. Given $\epsilon>0$ and access to the procedures $P_x$ and $P_y$ described above. Then the problem to output a vector $\beta={(\beta_1,\beta_2,...,\beta_d)}^T$ such that $\abs{\beta-\hat{\beta}}\le\epsilon$ and $\hat{\beta=\textbf{X}^\dagger\textbf{y}}$ can be solved by a gate-efficient quantum algorithm that makes $O(\frac{d^{2.5}\kappa^3}{\delta^2}{\rm poly}[{\rm log}_2(\frac{d\kappa}{\delta})])$ uses of $P_x$ and $P_y$, where $\delta=min\{1/d,\epsilon\}$.\\

\noindent This Quantum Linear Regression procedure is denoted as $QLR(d,\delta,\kappa)$. Then the predicted value vector $\hat{y}$ is calculated by the matrix multiplication
\begin{equation}
\hat{y}=X{\beta},
\end{equation}
and the residuals sequence is derived by a vector subtraction between the predicted values $\hat{y}$ and real values $y$:
\begin{equation}
u=y-\hat{y}.\
\end{equation} 
This should be a hybrid algorithm since classical algorithms can calculate matrix multiplications and subtractions with fewer restrictions and more efficiently.

Next, another  regression $QLR(L+1,\delta', \kappa')$ on time variable and lagged residuals will be employed to derive the statistical index. The lagged residuals $\Delta u_t$ is defined as the first-order difference and can be calculated efficiently by a vector subtraction:
\begin{equation}
\Delta u_t = u_t - u_{t-1}
\end{equation}
Then $QLR(L+1,\delta', \kappa')$ procedure shows:
\begin{equation}
\Delta u_t = \alpha+\beta t+\gamma u_{t-1}+\sum_{i=1}^{L-1} \delta_i\Delta u_{t-i}+\epsilon_t,
\end{equation}
where $L$ is the lag-length used in the ADF test, and $\beta$ is the coefficient of the time variable $t$. The test statistic $\text{DF}_T=\frac{\hat{\gamma}}{\text{SE}(\hat{\gamma})}$,
 where SE means standard error, can be computed by classcial computer more efficiently.

Finally, the result will be sent to be compared with a critical value table \cite{fuller1976introduction}. And the total algorithm is summarised as \textbf{Algorithm} \ref{alg:alg3}. \\

\begin{breakablealgorithm}
\caption{Quantum Cointegration Test Algorithm}
\label{alg:alg3}
\begin{algorithmic}[]
\Require ~~\\
$\kappa_0$:  the threshold for preselection\\
$T$:	the length of time interval\\
$J$:	the total number of stocks\\
$p_t^{(j)}$: the $j^{th}$ stock's price at time t.
\Ensure ~~\\
$(f,\beta)$flag and cointegrated coefficients.\\

\State Data Loading:
\State \indent$\ket{\psi_x}=\sum\limits_{t=0}^{T-1}\sum\limits_{j=0}^{J-1}r_t^{(j)}\ket{t}\ket{j}$:   amplitude encoding\\
\State Residual Construction Module: \\
\State \indent $QLR(d,\delta, \kappa)$ to derive $\beta$\\
\State \indent Classical matrix multiplication $\hat{y}=X{\beta}$\\
\State \indent Classical vector subtraction $\hat{u}=y-\hat{y}$\\
\State Statistics Calculation Module:\\
\State \indent Lagged residuals $\Delta u_t = u_t - u_{t-1}$\\
\State \indent $QLR(L+1,\delta', \kappa')$ to derive $\gamma$\\
\State \indent Classical test statistic $\text{DF}_T$\\
\State Comparison with Critical Value Table (\cite{fuller1976introduction})\\
\end{algorithmic}
\end{breakablealgorithm}
\noindent \\

\subsection*{B. Complexity Analysis}

In the following subsection, a detailed analysis of the algorithm's complexity is given. Suppose a single-round cointegration test on an $Nd$ design matrix where $N$ is the number of samples and $d$ is the number of variables. By lemma3, the regression coefficients can be derived directly with complexity to be $O(\frac{d^{2. 5}\kappa^3}{\delta^2}{\rm poly}(\log_2{\frac{d \kappa}{\delta}}))$.  Then the residuals can be computed directly in $O(Nd)$ steps. The result of this hybrid residual generation procedure is as follows:\\

\noindent\textbf{Lemma 5 (Complexity of Residuals Sequence Generation Procedure)} Suppose $X$ is an $N*d$ design matrix and $y$ the target vector, also we have $\epsilon$ the precision desired, and $\kappa$ is the condition number. Then the residuals sequence of regression can be derived with complexity $O(\frac{d^{2. 5}\kappa^3}{\delta^2}{\rm poly}(\log_2{\frac{d \kappa}{\delta}})+dN)$.\\

Besides this, it should be mentioned that an alternative method use \cite{schuld2016prediction}'s work to derive a predictor of a linear model. This method should be repeated $N$ times to derive the residuals sequence. Hence the total algorithm is $O(N\log N\kappa^2\epsilon^{-3})$.

\noindent 

Since the residuals derived from the above subroutine are intermediate instead of final results, it is important for us to analyse the error propagation of the cointegration test to control the global error:\\

\noindent\textbf{Lemma 6 (Bounded Error Propagation)} Suppose the error of the first regression(for residuals) be $\abs{\beta-\beta'}\le\epsilon$, then the error of the second regression(for cointegration test) is bounded by $\sqrt{L+2}\epsilon^2$ where $L$ is the lag length in the ADF test.
\\[8pt]
\noindent$Proof\ of\ \textbf{Lemma 6}. $
We can compute the error of residuals as follows:
Suppose that
\begin{equation}
u_t = X\beta-y
\end{equation}
and
\begin{equation}
u_t' = X\beta'-y
\end{equation}
are the residuals and estimated residuals, respectively. The error of the second regression variable $u_t$ is
\begin{align}
\abs{u_t-u_t'}&=\abs{(X\beta-y)-(X\beta'-y)}\\
&=\abs{X(\beta-\beta')}\\
&\le\epsilon,
\end{align}
Here (35) follows from $\lVert X \rVert=1$, and the errors of $\Delta u_t$ can be calculated as:
\begin{align}
\abs{\Delta u_t-\Delta u_t'}&=\abs{(u_t-u_{t-1}) - (u_t'-u_{t-1}')}\\
&\le \abs{u_t-u_{t-1}}+\abs{u_t'-u_{t-1}'}\\
&\le\epsilon+\epsilon=2\epsilon.
\end{align}

Regard these two error sequences as $2\epsilon$-bounded perturbation terms of the design matrix 
\begin{equation}
\hat{U} = U + E,
\end{equation}
in the second regression(30), by \cite{davies1975effect, beaton1976acceptability, stewart1977sensitivity}'s work the error propagation is bounded  as:
\begin{equation}
\lVert\gamma-\hat{\gamma}\rVert \le \sum f_{j}^2 \lVert\delta_j\rVert.
\end{equation}
Here $f_j=\sqrt{\gamma^2+\sum c_j {e_j}^2}$ is the sensitivity of the dpendence on the $j-th$ variable, and is bounded by $O(\sqrt{L+2}\epsilon)$. And $\delta_j$ is the error of $j-th$ term and hence is bounded by $O(\epsilon)$. Hence the total error propagation is bounded by $O(\sqrt{L+2}\epsilon^2)$.
\hfill$\blacksquare$\\

\noindent$Proof\ of\ \textbf{Theorem 3}. $With the facts above can the total complexity be calculated: the generation of the residuals will cost $O(\frac{d^{2. 5}\kappa^3}{\delta^2}{\rm poly}(\log_2{\frac{d \kappa}{\delta}})+dN)$; a second regression on residuals is implemented by $QLR$ again with propagated error $\epsilon' = \sqrt{L+2}\epsilon^2)$, condition number $\kappa'$ and $d=L+2$, and by lemma 4, the complexity is $O(\frac{{(L+2)}^{2. 5}\kappa'^3}{\delta'^2}{\rm poly}(\log_2{\frac{(L+2)\kappa'}{\delta'}}))$, where $\delta'=min\{1/(L+2),\sqrt{L+2}\epsilon^2\}$. The final complexity follows by a direct sum.
\hfill$\blacksquare$\\

\section{Realistic Case Analysis}
This section will analyze the quantum advantage of QSA in the realistic financial scenario of U.S. stock markets. There are mainly two kinds of characteristics data having significant influences on the algorithm complexity. One is the number of stocks: there are about 8000 stocks in the U.S. stock markets. Another is the trading time. The regular trading time of the New York Stock Exchange and the NASDAQ are both 6.5 hours per day. For the half-second time intervals aggregated quotes data, the length of data in one day is $N_0=6.5\times3600\times2=46800$. Furthermore there are about $l=253$ trading days one year on average. Hence the typical size of the time series data can be computed as
\begin{equation}
N=N_0\times l=46800\times253\approx1.2\times10^7.
\end{equation}

Under the cnosideration of the realistic case discussed above, there are mainly three reasons why QSA is more efficient than classical ones: First of all, in financial scenario, there are many different stocks, and it occupies only a tiny proportion of the searching space to find a multicollinearity portfolio out of thousands of stocks. The number of three-stock portfolios can exceed $M=C_{8000}^3\approx10^9$ while $M_0$, the number of multicointegrated pairs, is usually less than 1000. The proportion of non-multicollinear portfolios is estimated as
\begin{equation}
M_0/M\le10^{-6}.
\end{equation}
By (41), the classical benchmark is $O(N^2d)=10^{18}$, and the average complexity of our algorithm is mainly determined by the first preselection subroutine wtih complexity $O(\sqrt{d}\kappa_0^2\log{(1/\epsilon)}^2))={10}^8$(see details below). The primary reason for this acceleration is that the preselection procedure can search the multicollinearity without large matrix factorizations and regressions. Secondly, the problem size determined by sample number $N$ is supposed to be very large for our problem of high-frequency trading:  On the one hand, for high-frequency trading, there is a short time interval and a large number $N_0$ of trading date quotes of every single trading day. On the other hand, it does make sense in finance to consider a long time interval $l$ since it is a statistical arbitrage model instead of some models for prediction such as momentum trading. Finally, for the specific case of statistical arbitrage trading strategy, it is common and unavoidable to handle matrices with large condition number $\kappa$, resulting in high cost of computing resources and time complexity. Utilizing the ability to detect $\kappa$ by QCNCA, our algorithm is time variable one and adaptive to $\kappa$. Since most portfolios are with small $\kappa$ as discussed above, giving a bound $\kappa_0=1000$, our algorithm's complexity is about $O(\sqrt{d}\kappa_0^2\log{(1/\epsilon)}^2))=10^8$.

The number of qubits needed can be estimated as follows: According to the data size discussed above, the qubits needed to prepare for the initial state is about $\log{(1.2*10^7)}+\log{8000}\approx35$. The $QCNC(\kappa,\phi)$ circuit consists of simplified phase estimation subcircuits, and each subcircuit with 0.1 precision needs more than $4$ qubits. Moreover, the $VTPA$ circuit consists of $QCNC(\kappa_j,\phi)$ circuits for different $\kappa_j$, and hence more than 50 qubits are needed, which are hard for us to simulate. 

\section{CONCLUSION}

In this article, we introduce quantum algorithms for quantitative trading in the case of high-frequency statistical arbitrage and show the quantum advantage. Besides wxploring new financial applications, two heuristic algorithms are also developed as instruments: One is for the estimation of the condition number of a given matrix, which has not been considered and proposed before as far as we know. This algorithm can be applied to solve other problems where condition number is a primary influencing factor of the algorithm's complexity, such as quantum computational fluid dynamics and differential equation solution\cite{rebentrost2014quantum,berry2014high,berry2017quantum,childs2020high,clader2013preconditioned}. The other is the implemention of statistical cointegration test, which has many applications in time series, finance analysis.  Some modifications and exploration will be considered later to suit these exciting problems. 

During the analysis of QCNCA and VTPA's complexity, we provide a theoretical framework and show the quantum advantage under the assumption of uniform distribution. Since the real problems are complicated, many other statistical models and different distributions will be taken into consideration. By some modification in Eqs.(8-11), this method might still work  with different results of complexity, and this is our further research direction. Moreover, the work of circuit simplification and simulation will be done in the future.

\section*{ACKNOWLEDGEMENT}

This work was supported by the National Key Research and Development Program of China (Grant No. 2016YFA0301700), the National Natural Science Foundation of China (Grants Nos. 11625419), the Strategic Priority Research Program of the Chinese Academy of Sciences (Grant No. XDB24030600), and the Anhui Initiative in Quantum Information Technologies (Grants No. AHY080000).

\bibliography{exbib}

\end{document}